# A Study of the Orbital Periods of Deeply Eclipsing SW Sextantis Stars

David Boyd[1]

**Abstract**

Results are presented of a five-year project to study the orbital periods of eighteen deeply eclipsing novalike cataclysmic variables, collectively known as SW Sextantis stars, by combining new measurements of eclipse times with published measurements stretching back in some cases over fifty years. While the behaviour of many of these binary systems is consistent with a constant orbital period, it is evident that in several cases this is not true. Although the time span of these observations is relatively short, evidence is emerging that the orbital periods of some of these stars show cyclical variation with periods in the range 10–40 years. The two stars with the longest orbital periods, V363 Aur and BT Mon, also show secular period reduction with rates of $-6.6 \times 10^{-8}$ days/year and $-3.3 \times 10^{-8}$ days/year. New ephemerides are provided for all eighteen stars to facilitate observation of future eclipses.

## 1. SW Sex stars

SW Sex stars are an unofficial sub-class of cataclysmic variables (CVs), not in the General Catalogue of Variable Stars (GCVS; Samus et al. 2012), which was first proposed by Thorstensen et al. (1991) with the comment "…these objects show *mysterious* behaviour which is however highly *consistent* and *reproducible*." They are classified in the GCVS as novalike variables. The four prototype SW Sex stars were PX And, DW UMa, SW Sex, and V1315 Aql, which all appeared to share a common set of unusual properties (see below). Since then this class has expanded to include around fifty members of which about half are definite members and the others either probable or possible based on their observed characteristics. Don Hoard maintains an on-line list of SW Sex stars (Hoard et al. 2003).

These SW Sex stars have bright accretion disks, in some cases showing occasional VY Scl-type low states, but do not have the quasi-periodic outbursts seen in dwarf novae. They are often eclipsing systems with periods mostly in the range 3–4 hours. They may exhibit either positive or negative superhumps or both. Spectroscopically they show single-peaked Balmer and HeI emission lines, not double peaked lines as expected in high inclination CVs. Superimposed on the emission lines is a transient narrow absorption feature around phase 0.5. Phase offsets are observed between the radial velocity and eclipse ephemerides. Some systems exhibit modulated circular polarization indicating magnetic accretion onto the white dwarf. There is much variation in detail between individual systems and current models of SW Sex stars have difficulty explaining all of their observed properties. The general consensus seems to be that SW Sex stars contain accretion discs which are maintained in a bright state by a high, sustained mass-transfer rate and that these discs are complex in structure and may be variously eccentric, precessing, warped, tilted, or flared at the edge. The inner edge of the disc may also be truncated if the white dwarf is magnetic. For more information, see for example Hellier (1999), Gänsicke (2005), Rodriguez-Gil (2005), Rodriguez-Gil et al. (2007a), and Rodriguez-Gil et al. (2007b). Recently it has been claimed that the majority of novalike variables in the 3–4 hour orbital period range exhibit SW Sex-like properties to some extent (see Schmidtobreick et al. 2011). If true, this suggests that the SW Sex phenomenon may be a normal stage of CV evolution.

---

[1] 5 Silver Lane, West Challow, Wantage, OX12 9TX, UK; davidboyd@orion.me.uk

However, the bottom line at the moment seems to be that we really don't have a full understanding of the mechanisms which operate in SW Sex stars, and how they relate to other CVs with similar periods. But, as they appear to constitute the majority of CVs with orbital periods in the range 3–4 hours, they are important and need further study.

## 2. Aims of the project

This project was suggested to me in early 2007 by Boris Gänsicke at Warwick University who was interested to find out if studying eclipses of SW Sex stars would reveal evidence of changes in their orbital periods. Several of these stars had not been observed systematically for many years and were in need of new observations. The idea was therefore to combine published data on eclipse times going back in some cases over fifty years with new eclipse measurements to investigate the stability of their orbital periods.

The aims of the project were, for each star:

- to research all previously published eclipse times;
- to measure new eclipse times;
- to look for evidence of a change in orbital period;
- if found, to investigate its nature;
- to update ephemerides to aid future observations.

The eighteen SW Sex stars in Hoard's list which are deeply eclipsing, observable from the UK, and bright enough to yield accurate eclipse times with amateur-sized telescopes are the subject of this project. They are listed in Table 1 in order of increasing orbital period ($P_{orb}$) along with the numbers of eclipse times found in the literature and new measurements reported here.

## 3. Previously published eclipse times

Eclipse times of minimum of these stars were discovered in over twenty different publications. For each star a list of published eclipse times obtained photographically (PG), photoelectrically (PE), or using CCD cameras was assembled along with corresponding cycle (orbit) numbers. As far as possible all times were confirmed to be in Heliocentric Julian Date (HJD). A very small number of visual eclipse times were also found but after careful consideration it was decided not to use these in this analysis because of their significantly larger and generally unknown uncertainties. In total 740 published eclipse times were located for these eighteen stars. Limitation on space prevents listing previously published eclipse times here. These are available through the AAVSO ftp site at ftp://ftp.aavso.org/public/datasets/jboydd401.txt.

Many published eclipse times did not specify errors. By examining the scatter in eclipse times obtained photographically their error was estimated to be, on average, 0.005d and this value was assigned to all photographic times. For photoelectric and CCD measurements published without errors each published set of data was considered separately and the root-mean-square (rms) residual of all the times in that set calculated with respect to a locally fitted linear ephemeris. This value was then assigned as an error to all the times in that set. These errors were typically in the range 0.0004d to 0.001d. In cases where the errors quoted appeared to be unrealistically small, more realistic errors were estimated by the same method. Each published time of minimum was given a weight equal to the inverse square of its error.

## 4. New measurements of eclipse times

Eclipses were observed using either a 0.25-m or 0.35-m telescope, both equipped with Starlight Xpress SXV-H9 CCD cameras, located at West Challow Observatory near Oxford, UK. Image scales were 1.45 and 1.21 arcsec/pixel respectively. All measurements were made unfiltered for maximum photon statistics. Images were dark subtracted and flat fielded and a magnitude for the variable in each image derived with respect to between three and five nearby comparison stars using differential aperture photometry.

The dominant light source in these systems is the bright accretion disk, and its progressive eclipse by the secondary star results in eclipse profiles which are generally V-shaped with a rounded minimum. A quadratic fit was applied to the lower part of each eclipse from which the eclipse time of minimum and an associated analytical error were obtained. The magnitude at minimum was also obtained from this fit, enabling eclipse depths to be estimated. Some of these stars exhibit relatively large random fluctuations in light output which can persist during eclipses, indicating the source of these fluctuations has not been eclipsed. This can result in significant distortion of their eclipse profiles and consequently larger scatter in their measured times of minimum. In general it was found that the analytical errors from the quadratic fits underestimated the real scatter in eclipse times. By examining this scatter for each star over a short interval during which the eclipse times were varying linearly, a multiplying factor was found which was then applied to the analytical errors. For stars with the smoothest eclipses, a factor of 3 gave errors consistent with the scatter of eclipse times while for the most distorted eclipses a factor of 7 was required. Each measured time of minimum was given a weight equal to the inverse square of its associated error. All new times of minimum were converted to HJD. As shown in Table 1, 298 new eclipse times were measured, increasing the number of available eclipse times for these stars by 40%.

Initially a constant orbital period for each star was assumed and a linear ephemeris computed based only on published eclipse times. Predictions were then made of the expected times of future eclipses. Although in some cases these predictions were found to be inaccurate by up to an hour, in all cases it was possible to project the historical cycle count forward and unambiguously assign cycle numbers to new eclipses as they were observed.

For each star we now had the HJD of the time of minimum for every measured eclipse plus an error and a corresponding cycle number. New eclipse times measured for the eighteen stars in the project are listed in Table 2.

## 5. O–C analysis

For each star a constant orbital period was assumed and a weighted linear ephemeris was calculated based on all available eclipse times, both published and new. O–C (Observed minus Calculated) values for the time of each eclipse with respect to this linear ephemeris were calculated and an O–C diagram generated for each star. O–C values following the horizontal line at O–C = 0 would confirm that the orbital period was indeed constant. O–C values following an upwards curve would indicate that the period was increasing while a downwards curve would indicate that the period was decreasing. Sinusoidal behaviour would indicate that the orbital period was varying in a cyclical way, alternately increasing and decreasing.

## 6. Eclipsing SW Sex stars with orbital periods less than 4 hours

Most of the thirteen eclipsing SW Sex stars with orbital periods less than 4 hours have O–C diagrams which appear to be consistent with having a constant orbital period over the time span covered by the available observations, in some cases more than thirty years. However, in a few cases there is an indication of possible non-linear behaviour. This was investigated by applying a weighted sine fit to their O–C values using Period04 (Lenz and Breger 2005) and comparing the rms residuals of linear and sinusoidal ephemerides. The conclusion was that ten of the thirteen stars were consistent with having linear ephemerides and therefore a constant orbital period while three, SW Sex, LX Ser, and UU Aqr, gave at least 20% smaller rms residuals for sinusoidal ephemerides indicating possible cyclical variation of their orbital periods.

Linear ephemerides for these thirteen SW Sex stars are given in Table 3. These should provide an accurate basis for predicting the times of future eclipses. Table 4 lists the parameters of possible cyclical variation and rms residuals of sinusoidal and linear ephemerides for SW Sex, LX Ser, and UU Aqr.

Figure 1 shows O–C diagrams for the ten SW Sex stars with orbital periods less than 4 hours which are consistent with linear ephemerides. Previously published observations are marked as black dots and new eclipse times as red squares in this and subsequent figures. The larger scatter for some stars is primarily due to the less regular shape of their eclipses as noted above. Figure 2 shows O–C diagrams for SW Sex, LX Ser, and UU Aqr with dashed lines representing their sinusoidal ephemerides.

Given the length of their cyclical periods relative to the observed coverage and their relatively small amplitudes, more data are required to substantiate these cyclical interpretations. We do, however, note that similar behaviour has been recorded in several other eclipsing CVs, see for example Borges et al. (2008) and references therein.

## 7. Eclipsing SW Sex stars with orbital periods greater than 4 hours

Five of the SW Sex stars have orbital periods longer than 4 hours: RW Tri, 1RXS J064434.5+334451, AC Cnc, V363 Aur, and BT Mon. For all these stars the eclipse times appear, to varying degrees, to be inconsistent with the assumption of a constant orbital period. Each of these stars is now considered individually.

### 7.1. RW Tri

A total of 115 published and 21 new eclipse times are available for RW Tri starting in 1957. The O–C diagram for RW Tri representing the residuals to a linear ephemeris with long-term average orbital period 0.231883193(2) day is shown in Figure 3a. The scatter in the data is sufficiently large that a time calibration problem with some of the published times must be considered a possibility. We decided to exclude the 11 eclipse times around HJD 2449600 from subsequent analysis as their O–C values were more than 5 minutes larger than those before and after. Between approximately HJD 2442000 and HJD 2450000 the period slowly decreased. It then started to increase and is currently longer than the long-term average. Taken as a whole, the data suggest cyclical variation of the orbital period. A weighted sine fit to the O–C data using Period04 gives the results listed in Table 5 and shown as a dashed line in Figure 3a. Also listed in Table 5 are the rms residuals of sinusoidal and linear fits indicating that sinusoidal interpretation is statistically favoured. However, given the large scatter in the data and the fact that just over one possible cycle has been observed, a convincing analysis will require data over a much longer time span. An earlier analysis by Africano et al. (1978) suggested sinusoidal variation with a period of either 7.6 or 13.6 years but with the addition of more recent data neither of these periods survives.

A linear ephemeris fitted to the data over the past seven years which should be useful for predicting eclipses in the near future is given in Table 3.

The out-of-eclipse magnitude of RW Tri, including early measurements reported by Walker (1963), observations from the AAVSO International Database and from ASAS (Pojmański et al. 2005), and new observations reported here, is plotted in Figure 3b. This shows a slight brightening around HJD 2450000 but otherwise little change. Eclipse depth has remained approximately constant over the observed time span (Figure 3c). Table 6 lists measurements of eclipse depth for the five stars with long orbital periods.

### 7.2. 1RXS J064434.5+334451

Twenty unpublished eclipse times for 1RXS J064434.5+334451 from 2005 to 2008 were kindly provided by David Sing and Betsy Green, who first identified this star as a CV (Sing et al. 2007). The first times reported here were in 2010 and these were consistent with those of Sing and Green, giving the orbital period 0.26937447(4) day and the linear eclipse ephemeris given in Table 3.

Surprisingly, eclipses in March 2011 were about three minutes late relative to this ephemeris. This behaviour has since continued with most eclipses occurring between two and four minutes later than expected assuming a linear ephemeris based on observations up to and including 2010 (HJD < 2455500 – see Figure 4a). Since March 2011 the mean orbital period has been slightly shorter at 0.2693741(2) day. The out-of-eclipse magnitude experienced a rise prior to, and a dip following, the O–C discontinuity (Figure 4b). Eclipses became, temporarily, about 10% deeper after the O–C discontinuity (Figure 4c and Table 6).

### 7.3. AC Cnc

For AC Cnc, 46 published and 11 new eclipse times are available and fitting a linear ephemeris to these gives the O–C diagram shown in Figure 5a. Times measured before 1980 are photographic and have a large scatter. Using the more precise photoelectric and CCD measurements since 1980 (HJD > 2444000) gives an orbital period of 0.30047738(1) day and the linear eclipse ephemeris given in Table 3.

A recent paper by Qian et al. (2007) argues for a decreasing orbital period and also proposes a third body in the system causing sinusoidal modulation in the O–C diagram. A quadratic ephemeris calculated using the more reliable photoelectric and CCD measurements following HJD 2444000 and including the new times reported here gives a rate of period change of $2.1(2.2) \times 10^{-9}$ days/year, considerably smaller than the rate of $12.4(4.4) \times 10^{-9}$ days/year found by Qian et al. and consistent with no secular period change. However, there is an indication of cyclical behaviour in the O–C diagram in Figure 5a. A weighted sine fit to the O–C data after HJD 2444000 gives the results listed in Table 5 and shown as a dashed line in Figure 5a. This is a shorter period and smaller amplitude than proposed by Qian et al. Observations over a much longer period are required to establish the reality and true parameters of this modulation. With few measurements available, there is little indication of variation in either the out-of-eclipse magnitude of AC Cnc (Figure 5b) or the eclipse depth (Figure 5c and Table 6).

### 7.4. V363 Aur (also known as Lanning 10)

The data for V363 Aur comprise 18 published and 27 new eclipse times and show a significant curvature in the O–C diagram with respect to a linear ephemeris, indicating a reducing orbital period (Figure 6a). A quadratic ephemeris, shown as a dashed line in Figure 6a, gives a mean rate of period change of $dP/dt = -6.6(2) \times 10^{-8}$ days/year over the 31 years covered by the data. The O–C residuals to this quadratic ephemeris (Figure 6b) show an apparently cyclical variation. A weighted sine fit to the residuals of the quadratic ephemeris gives the results listed in Table 5 and shown as a dashed line in Figure 6b, but these results must be considered speculative as only one cycle has been observed.

Over the past six years the mean orbital period has been 0.32124073(3) day and the eclipse times are well fitted by the linear ephemeris given in Table 3.

Figure 6c shows the out-of-eclipse magnitude of V363 Aur obtained from the AAVSO database plus our new measurements. Although the scatter is large, there appears to have been a slight dip centred around HJD 2449000. Figure 6d and Table 6 show the depth of eclipses of V363 Aur over the same interval. Although there are little data in the early years, recently there has been a progressive reduction in eclipse depth.

### 7.5. BT Mon

BT Mon is the progenitor system of a classical nova outburst observed in 1939. There are 8 published eclipse times plus 14 new times covering a 34 year time span. An O–C diagram with respect to a linear ephemeris shows significant curvature indicating a reducing orbital period (Figure 7a). A quadratic ephemeris, shown as a dashed line in Figure 7a, gives a mean rate of period change of $dP/dt = -3.3(2) \times 10^{-8}$ days/year. The O–C residuals to this quadratic ephemeris are shown in Figure 7b along with a dashed line indicating the results of a weighted sine fit whose parameters are listed in Table 5. This sinusoidal ephemeris is marginally favoured over the quadratic ephemeris although, as before, this conclusion must remain tentative until more data are available.

Published eclipse times are scarce in the middle of this time span. Magnitude measurements of BT Mon obtained with the Roboscope system (Honeycutt 2003) between 1991 and 2005 were kindly provided by Kent Honeycutt. The Roboscope data were divided into two groups, before and after the start of 1999. By adopting mean orbital periods from the above quadratic ephemeris for each of these time intervals, two additional eclipse times have been synthesised using the Roboscope data. These have larger errors than directly measured eclipse times and are shown as green triangles in Figures 7. They were not used in the above analysis but are consistent with its results and slightly favour the sinusoidal interpretation.

Over the last seventeen years the mean orbital period has been 0.33381322(2) day and eclipse times in the near future may be represented by the linear ephemeris given in Table 3.

Plotting out-of-eclipse magnitudes from Roboscope together with our new data (Figure 7c), we see a noticeable dip around HJD 2451000 followed by a gradual increase. Although there are little data, the eclipse depth shows a slowly decreasing trend over the same interval (Figure 7d and Table 6).

### 8. Conclusions

When this project started, most published analyses of SW Sex stars concluded that they had constant orbital periods. While the new data confirm that this is true for many of these stars, for some it is clearly not the case. There is a significant difference between the behaviour of stars with orbital periods below and above 4 hours. Below 4 hours, ten of the thirteen stars appear to have constant orbital periods with three showing possible signs of low amplitude cyclical variation. The longer period stars all show more dynamic behaviour with either a sudden change of orbital period or larger amplitude cyclical variation, either with or without a secular period change.

Eclipse times for all these stars will continue to be monitored to see if those with constant periods maintain this behaviour and in the other more interesting cases with longer orbital periods to discover what light further data will shed on the tentative interpretations presented here.

## 9. Acknowledgements

I am grateful to Boris Gänsicke for suggesting this project and for his continuing support and encouragement. I am indebted to David Sing and Betsy Green for providing unpublished data on 1RXS J064434.5+334451 and to Kent Honeycutt for providing unpublished Roboscope data on BT Mon. I acknowledge with thanks that this research has made use of variable star observations from the AAVSO International Database contributed by researchers worldwide, data from the All Sky Automated Survey, and from NASA's Astrophysics Data System. Helpful comments from an anonymous referee have improved the paper.

Table 1. Eclipsing SW Sex stars studied in this project.

| Name | $P_{orb}$ (hrs) | Previously published eclipse times | New eclipse times reported here |
|---|---|---|---|
| HS 0728+6738 | 3.21 | 14 | 24 |
| SW Sex | 3.24 | 32 | 11 |
| DW UMa | 3.28 | 176 | 20 |
| HS 0129+2933 = TT Tri | 3.35 | 27 | 11 |
| V1315 Aql | 3.35 | 71 | 16 |
| PX And | 3.51 | 38 | 22 |
| HS 0455+8315 | 3.57 | 5 | 15 |
| HS 0220+0603 | 3.58 | 13 | 13 |
| BP Lyn | 3.67 | 16 | 13 |
| BH Lyn | 3.74 | 29 | 16 |
| LX Ser | 3.80 | 50 | 10 |
| UU Aqr | 3.93 | 50 | 15 |
| V1776 Cyg | 3.95 | 12 | 17 |
| RW Tri | 5.57 | 115 | 21 |
| 1RXS J064434.5+334451 | 6.47 | 20 | 22 |
| AC Cnc | 7.21 | 46 | 11 |
| V363 Aur = Lanning 10 | 7.71 | 18 | 27 |
| BT Mon | 8.01 | 8 | 14 |
| Total | | 740 | 298 |

Table 2. Eclipse times for stars measured in this project with errors and corresponding cycle numbers.

| Eclipse time of minimum (HJD) | Error (d) | Cycle number | Eclipse time of minimum (HJD) | Error (d) | Cycle number |
|---|---|---|---|---|---|
| **HS 0728+6738** | | | 2454185.43702 | 0.00044 | 72965 |
| 2453810.40077 | 0.00041 | 13539 | 2454186.38145 | 0.00029 | 72972 |
| 2453836.45653 | 0.00024 | 13734 | 2454553.41407 | 0.00048 | 75692 |
| 2453851.42254 | 0.00023 | 13846 | 2454564.34410 | 0.00020 | 75773 |
| 2453853.42648 | 0.00013 | 13861 | 2454906.41325 | 0.00019 | 78308 |
| 2454174.51418 | 0.00022 | 16264 | 2454907.49269 | 0.00019 | 78316 |
| 2454181.32859 | 0.00025 | 16315 | 2455260.35696 | 0.00018 | 80931 |
| 2454185.33706 | 0.00025 | 16345 | 2455278.43821 | 0.00012 | 81065 |
| 2454186.40643 | 0.00024 | 16353 | 2455630.35814 | 0.00026 | 83673 |
| 2454473.42029 | 0.00023 | 18501 | 2455660.44910 | 0.00014 | 83896 |
| 2454493.33001 | 0.00023 | 18650 | 2455662.33853 | 0.00028 | 83910 |
| 2454507.35967 | 0.00039 | 18755 | **DW UMa** | | |
| 2454835.39541 | 0.00032 | 21210 | 2454181.41978 | 0.00019 | 58214 |
| 2454891.38182 | 0.00010 | 21629 | 2454185.38111 | 0.00030 | 58243 |
| 2454895.39084 | 0.00009 | 21659 | 2454224.45051 | 0.00044 | 58529 |
| 2454907.41644 | 0.00022 | 21749 | 2454473.34780 | 0.00038 | 60351 |
| 2455188.41832 | 0.00021 | 23852 | 2454564.46466 | 0.00020 | 61018 |
| 2455191.35834 | 0.00014 | 23874 | 2454580.44785 | 0.00033 | 61135 |
| 2455200.31029 | 0.00019 | 23941 | 2454580.58433 | 0.00027 | 61136 |
| 2455515.38459 | 0.00024 | 26299 | 2454588.37104 | 0.00029 | 61193 |
| 2455520.32865 | 0.00038 | 26336 | 2454588.50711 | 0.00019 | 61194 |
| 2455533.42346 | 0.00028 | 26434 | 2454593.42488 | 0.00022 | 61230 |
| 2455889.38551 | 0.00036 | 29098 | 2454596.43092 | 0.00034 | 61252 |
| 2455891.39036 | 0.00024 | 29113 | 2454884.39723 | 0.00025 | 63360 |
| 2455893.39432 | 0.00019 | 29128 | 2454892.32009 | 0.00025 | 63418 |
| **SW Sex** | | | 2455239.30026 | 0.00022 | 65958 |

| Eclipse time of minimum (HJD) | Error (d) | Cycle number | Eclipse time of minimum (HJD) | Error (d) | Cycle number |
|---|---|---|---|---|---|
| 2455263.34322 | 0.00015 | 66134 | **HS 0455+8315** | | |
| 2455270.31000 | 0.00014 | 66185 | 2454061.40139 | 0.00016 | 14807 |
| 2455278.37037 | 0.00017 | 66244 | 2454063.48351 | 0.00020 | 14821 |
| 2455627.39978 | 0.00017 | 68799 | 2454078.35643 | 0.00014 | 14921 |
| 2455628.35604 | 0.00020 | 68806 | 2454112.41335 | 0.00017 | 15150 |
| 2455629.31205 | 0.00030 | 68813 | 2454114.49593 | 0.00023 | 15164 |
| **HS 0129+2933** | | | 2454115.38831 | 0.00017 | 15170 |
| 2454061.46332 | 0.00014 | 10892 | 2454895.44552 | 0.00018 | 20415 |
| 2454081.29219 | 0.00016 | 11034 | 2454906.45070 | 0.00013 | 20489 |
| 2454086.45848 | 0.00008 | 11071 | 2454907.34318 | 0.00026 | 20495 |
| 2455106.37036 | 0.00038 | 18375 | 2455065.43666 | 0.00029 | 21558 |
| 2455188.47729 | 0.00030 | 18963 | 2455495.39753 | 0.00032 | 24449 |
| 2455191.27007 | 0.00019 | 18983 | 2455519.49112 | 0.00017 | 24611 |
| 2455460.49099 | 0.00013 | 20911 | 2455526.48082 | 0.00018 | 24658 |
| 2455533.38206 | 0.00022 | 21433 | 2455835.38030 | 0.00021 | 26735 |
| 2455827.45860 | 0.00014 | 23539 | 2455850.40114 | 0.00018 | 26836 |
| 2455835.41776 | 0.00016 | 23596 | **HS 0220+0603** | | |
| 2455836.39518 | 0.00010 | 23603 | 2454061.32109 | 0.00048 | 10038 |
| **V1315 Aql** | | | 2454081.31479 | 0.00032 | 10172 |
| 2454272.50437 | 0.00018 | 59916 | 2454081.46403 | 0.00018 | 10173 |
| 2454306.44865 | 0.00027 | 60159 | 2454086.38783 | 0.00026 | 10206 |
| 2454313.43262 | 0.00072 | 60209 | 2455156.35608 | 0.00028 | 17377 |
| 2454651.48330 | 0.00048 | 62629 | 2455188.43603 | 0.00027 | 17592 |
| 2454670.48100 | 0.00046 | 62765 | 2455200.37262 | 0.00034 | 17672 |
| 2454810.31097 | 0.00082 | 63766 | 2455490.43180 | 0.00028 | 19616 |
| 2455004.47952 | 0.00029 | 65156 | 2455495.35697 | 0.00031 | 19649 |
| 2455006.43480 | 0.00049 | 65170 | 2455515.34977 | 0.00031 | 19783 |
| 2455038.42351 | 0.00055 | 65399 | 2455533.40410 | 0.00029 | 19904 |
| 2455052.39293 | 0.00070 | 65499 | 2455867.48013 | 0.00024 | 22143 |
| 2455463.36184 | 0.00047 | 68441 | 2455884.48964 | 0.00012 | 22257 |
| 2455464.33978 | 0.00036 | 68448 | **BP Lyn** | | |
| 2455490.32143 | 0.00026 | 68634 | 2454186.44462 | 0.00069 | 41257 |
| 2455777.38468 | 0.00040 | 70689 | 2454891.36892 | 0.00095 | 45870 |
| 2455783.39087 | 0.00040 | 70732 | 2454906.49781 | 0.00084 | 45969 |
| 2455903.24546 | 0.00047 | 71590 | 2455239.32473 | 0.00058 | 48147 |
| **PX And** | | | 2455260.41122 | 0.00042 | 48285 |
| 2454318.44729 | 0.00051 | 34708 | 2455263.31415 | 0.00049 | 48304 |
| 2454319.47234 | 0.00046 | 34715 | 2455571.38461 | 0.00074 | 50320 |
| 2454325.47261 | 0.00036 | 34756 | 2455594.30701 | 0.00042 | 50470 |
| 2454448.40773 | 0.00061 | 35596 | 2455619.52087 | 0.00059 | 50635 |
| 2454473.28943 | 0.00051 | 35766 | 2455914.44759 | 0.00041 | 52565 |
| 2454503.29163 | 0.00022 | 35971 | 2455930.34125 | 0.00063 | 52669 |
| 2454761.45718 | 0.00049 | 37735 | 2455932.32762 | 0.00066 | 52682 |
| 2454770.38547 | 0.00069 | 37796 | 2455942.41314 | 0.00039 | 52748 |
| 2455064.40680 | 0.00108 | 39805 | **BH Lyn** | | |
| 2455066.45577 | 0.00069 | 39819 | 2454181.48914 | 0.00029 | 44915 |
| 2455173.29503 | 0.00032 | 40549 | 2454186.32132 | 0.00042 | 44946 |
| 2455186.32065 | 0.00020 | 40638 | 2454199.41436 | 0.00053 | 45030 |
| 2455188.36884 | 0.00125 | 40652 | 2454482.32954 | 0.00048 | 46845 |
| 2455191.29553 | 0.00055 | 40672 | 2454834.45234 | 0.00046 | 49104 |
| 2455201.24653 | 0.00014 | 40740 | 2454884.33284 | 0.00052 | 49424 |
| 2455460.43876 | 0.00028 | 42511 | 2455247.36666 | 0.00027 | 51753 |
| 2455495.26963 | 0.00061 | 42749 | 2455260.46000 | 0.00033 | 51837 |
| 2455515.46733 | 0.00025 | 42887 | 2455267.31793 | 0.00059 | 51881 |
| 2455795.43984 | 0.00024 | 44800 | 2455594.34608 | 0.00035 | 53979 |
| 2455819.44115 | 0.00069 | 44964 | 2455628.32676 | 0.00041 | 54197 |
| 2455823.39248 | 0.00044 | 44991 | 2455670.41251 | 0.00040 | 54467 |
| 2455901.25250 | 0.00064 | 45523 | 2455675.40111 | 0.00031 | 54499 |

| Eclipse time of minimum (HJD) | Error (d) | Cycle number | Eclipse time of minimum (HJD) | Error (d) | Cycle number |
|---|---|---|---|---|---|
| 2455895.34197 | 0.00038 | 55910 | 2455487.34152 | 0.00042 | 61919 |
| 2455902.35570 | 0.00039 | 55955 | 2455490.35562 | 0.00017 | 61932 |
| 2455941.32605 | 0.00040 | 56205 | 2455533.48590 | 0.00023 | 62118 |
| **LX Ser** | | | 2455822.41233 | 0.00026 | 63364 |
| 2454316.41420 | 0.00032 | 63266 | 2455828.44141 | 0.00023 | 63390 |
| 2454628.52570 | 0.00023 | 65236 | 2455867.39741 | 0.00048 | 63558 |
| 2454976.44297 | 0.00038 | 67432 | 2455881.31079 | 0.00014 | 63618 |
| 2454994.50414 | 0.00026 | 67546 | 2455889.42621 | 0.00028 | 63653 |
| 2455001.47525 | 0.00033 | 67590 | 2455914.23796 | 0.00028 | 63760 |
| 2455037.43960 | 0.00020 | 67817 | 2455950.41154 | 0.00024 | 63916 |
| 2455662.45627 | 0.00040 | 71762 | 2455953.42610 | 0.00051 | 63929 |
| 2455663.40637 | 0.00045 | 71768 | 2455957.36910 | 0.00018 | 63946 |
| 2455672.43730 | 0.00041 | 71825 | **1RXS J064434.5+334451** | | |
| 2455778.42860 | 0.00031 | 72494 | 2455307.42924 | 0.00074 | 7067 |
| **UU Aqr** | | | 2455310.39210 | 0.00056 | 7078 |
| 2454323.44995 | 0.00046 | 48760 | 2455313.35557 | 0.00049 | 7089 |
| 2454357.47405 | 0.00027 | 48968 | 2455627.44814 | 0.00048 | 8255 |
| 2454365.48955 | 0.00036 | 49017 | 2455629.33392 | 0.00043 | 8262 |
| 2454728.47437 | 0.00051 | 51236 | 2455634.45149 | 0.00035 | 8281 |
| 2454735.34486 | 0.00034 | 51278 | 2455655.46296 | 0.00045 | 8359 |
| 2454736.32601 | 0.00056 | 51284 | 2455658.42635 | 0.00025 | 8370 |
| 2454789.32574 | 0.00032 | 51608 | 2455682.39947 | 0.00042 | 8459 |
| 2455038.45994 | 0.00069 | 53131 | 2455685.36351 | 0.00051 | 8470 |
| 2455059.39716 | 0.00052 | 53259 | 2455850.48993 | 0.00045 | 9083 |
| 2455106.34585 | 0.00043 | 53546 | 2455854.53082 | 0.00023 | 9098 |
| 2455469.49424 | 0.00052 | 55766 | 2455891.43482 | 0.00015 | 9235 |
| 2455490.26865 | 0.00048 | 55893 | 2455905.44214 | 0.00063 | 9287 |
| 2455778.49715 | 0.00019 | 57655 | 2455914.33106 | 0.00043 | 9320 |
| 2455795.50952 | 0.00019 | 57759 | 2455924.29847 | 0.00046 | 9357 |
| 2455893.33048 | 0.00019 | 58357 | 2455932.37955 | 0.00032 | 9387 |
| **V1776 Cyg** | | | 2455949.35041 | 0.00027 | 9450 |
| 2454238.48406 | 0.00059 | 45659 | 2455953.38926 | 0.00037 | 9465 |
| 2454254.46252 | 0.00044 | 45756 | 2455957.43085 | 0.00052 | 9480 |
| 2454306.51977 | 0.00050 | 46072 | 2455959.31737 | 0.00024 | 9487 |
| 2454314.42730 | 0.00053 | 46120 | 2455960.39430 | 0.00028 | 9491 |
| 2454646.54029 | 0.00092 | 48136 | **AC Cnc** | | |
| 2454668.44971 | 0.00092 | 48269 | 2454199.45197 | 0.00026 | 32978 |
| 2454670.42804 | 0.00080 | 48281 | 2454507.44198 | 0.00021 | 34003 |
| 2454770.42363 | 0.00115 | 48888 | 2454891.45161 | 0.00036 | 35281 |
| 2454994.46940 | 0.00068 | 50248 | 2454892.35306 | 0.00032 | 35284 |
| 2455037.46488 | 0.00052 | 50509 | 2455260.43835 | 0.00023 | 36509 |
| 2455057.39969 | 0.00051 | 50630 | 2455270.35440 | 0.00042 | 36542 |
| 2455176.34096 | 0.00062 | 51352 | 2455619.50814 | 0.00082 | 37704 |
| 2455460.34923 | 0.00100 | 53076 | 2455630.32565 | 0.00024 | 37740 |
| 2455494.45030 | 0.00101 | 53283 | 2455675.39674 | 0.00047 | 37890 |
| 2455778.46040 | 0.00052 | 55007 | 2455949.43118 | 0.00029 | 38802 |
| 2455849.46194 | 0.00052 | 55438 | 2455959.34723 | 0.00034 | 38835 |
| 2455893.28160 | 0.00088 | 55704 | **V363 Aur** | | |
| **RW Tri** | | | 2454181.39163 | 0.00043 | 29957 |
| 2454392.38737 | 0.00024 | 57197 | 2454392.44674 | 0.00017 | 30614 |
| 2454419.51756 | 0.00027 | 57314 | 2454447.37885 | 0.00024 | 30785 |
| 2454447.34346 | 0.00020 | 57434 | 2454471.47221 | 0.00031 | 30860 |
| 2454789.37226 | 0.00041 | 58909 | 2454473.39980 | 0.00037 | 30866 |
| 2454810.47333 | 0.00064 | 59000 | 2454810.38137 | 0.00031 | 31915 |
| 2454835.28542 | 0.00050 | 59107 | 2454827.40653 | 0.00042 | 31968 |
| 2455063.45767 | 0.00047 | 60091 | 2454835.43772 | 0.00044 | 31993 |
| 2455106.35664 | 0.00047 | 60276 | 2454891.33360 | 0.00054 | 32167 |
| 2455172.44338 | 0.00026 | 60561 | 2454892.29747 | 0.00021 | 32170 |

| Eclipse time of minimum (HJD) | Error (d) | Cycle number |
|---|---|---|
| 2455188.48144 | 0.00054 | 33092 |
| 2455191.37255 | 0.00040 | 33101 |
| 2455200.36736 | 0.00026 | 33129 |
| 2455515.50429 | 0.00013 | 34110 |
| 2455516.46885 | 0.00021 | 34113 |
| 2455524.49896 | 0.00034 | 34138 |
| 2455526.42586 | 0.00026 | 34144 |
| 2455627.29626 | 0.00020 | 34458 |
| 2455634.36298 | 0.00020 | 34480 |
| 2455649.46157 | 0.00047 | 34527 |
| 2455854.41351 | 0.00026 | 35165 |
| 2455888.46463 | 0.00016 | 35271 |
| 2455891.35618 | 0.00021 | 35280 |
| 2455905.49122 | 0.00039 | 35324 |
| 2455914.48560 | 0.00015 | 35352 |
| 2455950.46438 | 0.00013 | 35464 |
| 2455954.31900 | 0.00028 | 35476 |
| **BT Mon** | | |
| 2454447.47617 | 0.00043 | 32820 |
| 2454891.44778 | 0.00052 | 34150 |
| 2454892.44988 | 0.00045 | 34153 |
| 2455238.27878 | 0.00050 | 35189 |
| 2455239.28089 | 0.00082 | 35192 |
| 2455257.30609 | 0.00035 | 35246 |
| 2455260.31093 | 0.00041 | 35255 |
| 2455277.33531 | 0.00068 | 35306 |
| 2455571.42510 | 0.00058 | 36187 |
| 2455595.46030 | 0.00062 | 36259 |
| 2455600.46698 | 0.00093 | 36274 |
| 2455619.49354 | 0.00048 | 36331 |
| 2455960.31808 | 0.00089 | 37352 |
| 2455968.33013 | 0.00047 | 37376 |

Table 3. Linear ephemerides for the SW Sex stars in the project. For RW Tri, 1RXS J064434.5+334451, AC Cnc, V363 Aur, and BT Mon this linear ephemeris only represents behaviour in the recent past. Over longer time intervals their behaviour is more complex (see text).

| HS 0728+6738 | 2452001.32739(8) + 0.133619437(4) * E |
|---|---|
| SW Sex | 2444339.64968(11) + 0.134938490(2) * E |
| DW UMa | 2446229.00601(8) + 0.136606547(2) * E |
| HS 0129+2933 = TT Tri | 2452540.53218(9) + 0.139637462(6) * E |
| V1315 Aql | 2445902.84037(10) + 0.139689961(2) * E |
| PX And | 2449238.83661(17) + 0.146352746(4) * E |
| HS 0455+8315 | 2451859.24679(15) + 0.148723901(8) * E |
| HS 0220+0603 | 2452563.57407(7) + 0.149207696(5) * E |
| BP Lyn | 2447881.85799(23) + 0.152812531(6) * E |
| BH Lyn | 2447180.33522(41) + 0.155875629(8) * E |
| LX Ser | 2444293.02345(18) + 0.158432492(3) * E |
| UU Aqr | 2446347.26651(6) + 0.163580450(2) * E |
| V1776 Cyg | 2446716.67956(27) + 0.164738679(6) * E |
| RW Tri | 2441129.35318(49) + 0.231883392(9) * E |
| 1RXS J064434.5+334451 | 2453403.75955(12) + 0.26937447(4) * E |
| AC Cnc | 2444290.30892(36) + 0.30047738(1) * E |
| V363 Aur = Lanning 10 | 2444557.98318(89) + 0.32124073(3) * E |
| BT Mon | 2443491.72616(45) + 0.33381322(2) * E |

Table 4. Parameters of possible cyclical variation in orbital period for SW Sex, LX Ser, and UU Aqr.

|  | Cyclical period (year) | Semi-amplitude (sec) | Sinusoidal ephemeris rms residual | Linear ephemeris rms residual |
|---|---|---|---|---|
| SW Sex | 24.0(7) | 69(5) | 32.1 | 65.2 |
| LX Ser | 28(2) | 48(6) | 55.7 | 69.4 |
| UU Aqr | 20.3(6) | 48(4) | 34.9 | 43.6 |

Table 5. Parameters of possible cyclical variation in orbital period for RW Tri, AC Cnc, V363 Aur, and BT Mon (* only including data after HJD 2444000).

|  | Cyclical period (year) | Semi-amplitude (sec) | Sinusoidal ephemeris rms residual | Linear ephemeris rms residual | Quadratic ephemeris rms residual |
|---|---|---|---|---|---|
| RW Tri | 36.7(4) | 161(5) | 80.8 | 128.0 |  |
| AC Cnc* | 13.5(3) | 140(13) | 106.8 | 141.3 | 139.5 |
| V363 Aur | 27.7(7) | 119(6) | 58.8 |  | 92.2 |
| BT Mon | 29(2) | 113(15) | 62.0 |  | 67.7 |

Table 6. Eclipse depth measured in this project for RW Tri, 1RXS J064434.5+334451, AC Cnc, V363 Aur, and BT Mon.

| Eclipse time (HJD) | Eclipse depth (mag) | Eclipse time (HJD) | Eclipse depth (mag) |
|---|---|---|---|
| **RW Tri** |  | **AC Cnc** |  |
| 2454392.38737 | 1.72 | 2454199.45197 | 0.96 |
| 2454419.51756 | 1.84 | 2454507.44198 | 1.04 |
| 2454447.34346 | 1.96 | 2454891.45161 | 0.94 |
| 2454810.47333 | 1.84 | 2454892.35306 | 0.92 |
| 2454835.28542 | 1.63 | 2455260.43835 | 1.00 |
| 2455063.45767 | 1.76 | 2455630.32565 | 0.87 |
| 2455106.35664 | 1.92 | 2455949.43118 | 1.14 |
| 2455172.44338 | 1.89 | **V363 Aur** |  |
| 2455487.34152 | 1.78 | 2454392.44674 | 0.80 |
| 2455490.35562 | 1.71 | 2454447.37885 | 0.71 |
| 2455533.48590 | 1.84 | 2454471.47221 | 0.90 |
| 2455822.41233 | 1.62 | 2454473.39980 | 0.83 |
| 2455828.44141 | 1.43 | 2454810.38137 | 0.62 |
| 2455867.39741 | 1.59 | 2454827.40653 | 0.66 |
| 2455881.31079 | 1.81 | 2454835.43772 | 0.77 |
| 2455889.42621 | 1.89 | 2454892.29747 | 0.64 |
| 2455914.23796 | 1.69 | 2455191.37255 | 0.76 |
| 2455950.41154 | 2.06 | 2455515.50429 | 0.56 |
| 2455953.42610 | 1.97 | 2455516.46885 | 0.68 |
| 2455957.36910 | 1.56 | 2455524.49896 | 0.60 |
| **1RXS J064434.5+334451** |  | 2455526.42586 | 0.48 |
| 2455307.42924 | 1.13 | 2455627.29626 | 0.62 |
| 2455310.39210 | 1.06 | 2455634.36298 | 0.68 |
| 2455627.44814 | 1.26 | 2455649.46157 | 0.69 |
| 2455629.33392 | 1.27 | 2455854.41351 | 0.58 |
| 2455634.45149 | 0.90 | 2455888.46463 | 0.53 |
| 2455655.46296 | 1.22 | 2455891.35618 | 0.62 |
| 2455658.42635 | 1.29 | 2455905.49122 | 0.65 |
| 2455682.39947 | 1.31 | 2455950.46438 | 0.71 |
| 2455685.36351 | 1.23 | **BT Mon** |  |
| 2455850.48993 | 1.17 | 2454891.44778 | 1.74 |
| 2455854.53082 | 1.11 | 2454892.44988 | 1.82 |
| 2455891.43482 | 1.12 | 2455257.30609 | 2.01 |
| 2455905.44214 | 1.13 | 2455260.31093 | 1.90 |
| 2455914.33106 | 1.08 | 2455277.33531 | 1.96 |
| 2455932.37955 | 1.14 | 2455571.42510 | 1.61 |
| 2455949.35041 | 1.02 | 2455960.31808 | 1.49 |
| 2455953.38926 | 1.07 | 2455968.33013 | 1.62 |
| 2455957.43085 | 0.94 |  |  |
| 2455959.31737 | 0.97 |  |  |
| 2455960.39430 | 1.14 |  |  |

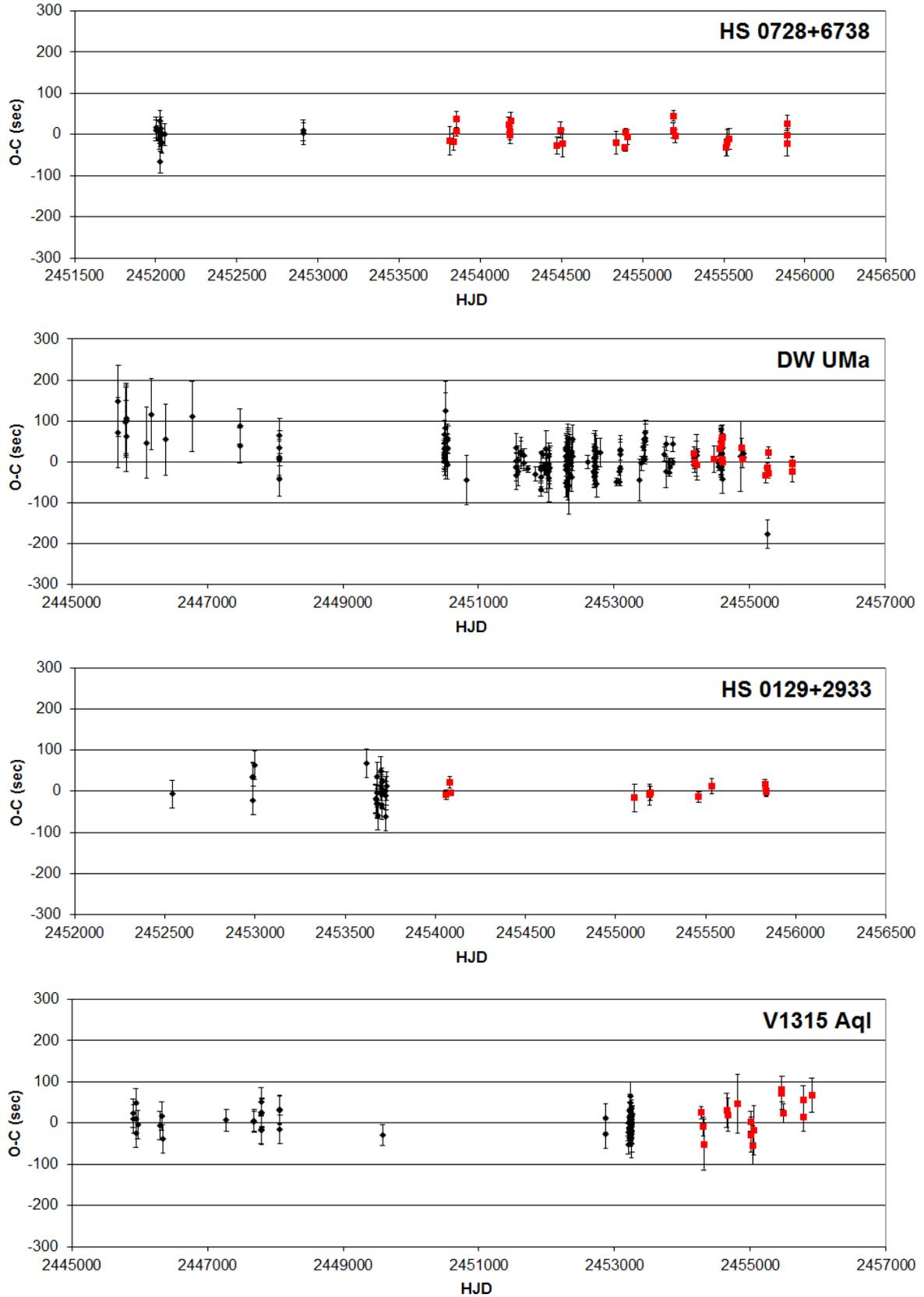

Figure 1. O–C diagrams with respect to the linear ephemerides in Table 3 for those SW Sex stars with Porb < 4 hours which are consistent with constant orbital periods. Previously published observations are marked as black dots and new eclipse times as red squares in this and subsequent figures (continued on next page).

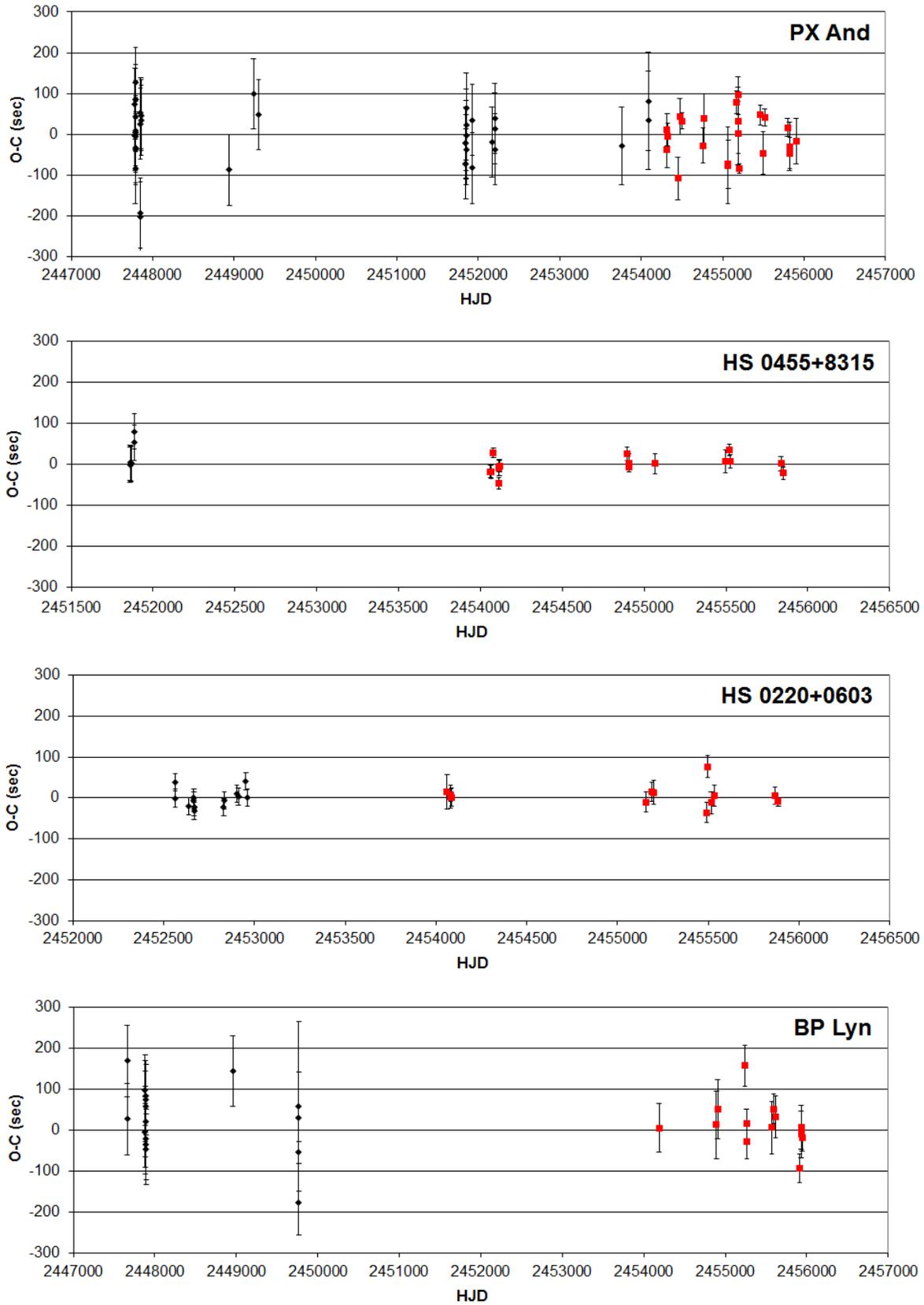

Figure 1. O–C diagrams with respect to the linear ephemerides in Table 3 for those SW Sex stars with Porb < 4 hours which are consistent with constant orbital periods. Previously published observations are marked as black dots and new eclipse times as red squares in this and subsequent figures (continued on next page).

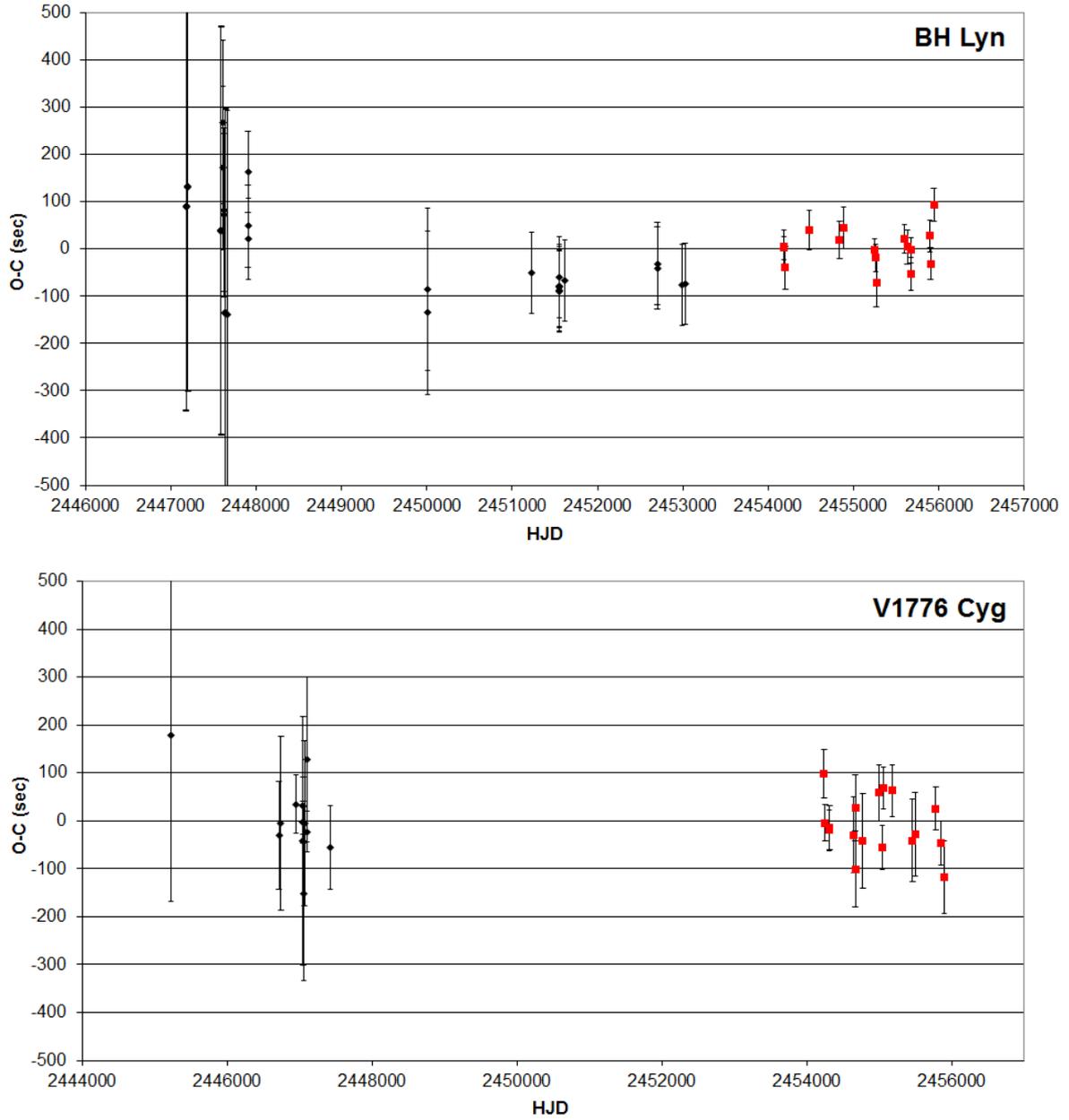

Figure 1. O–C diagrams with respect to the linear ephemerides in Table 3 for those SW Sex stars with Porb < 4 hours which are consistent with constant orbital periods. Previously published observations are marked as black dots and new eclipse times as red squares in this and subsequent figures.

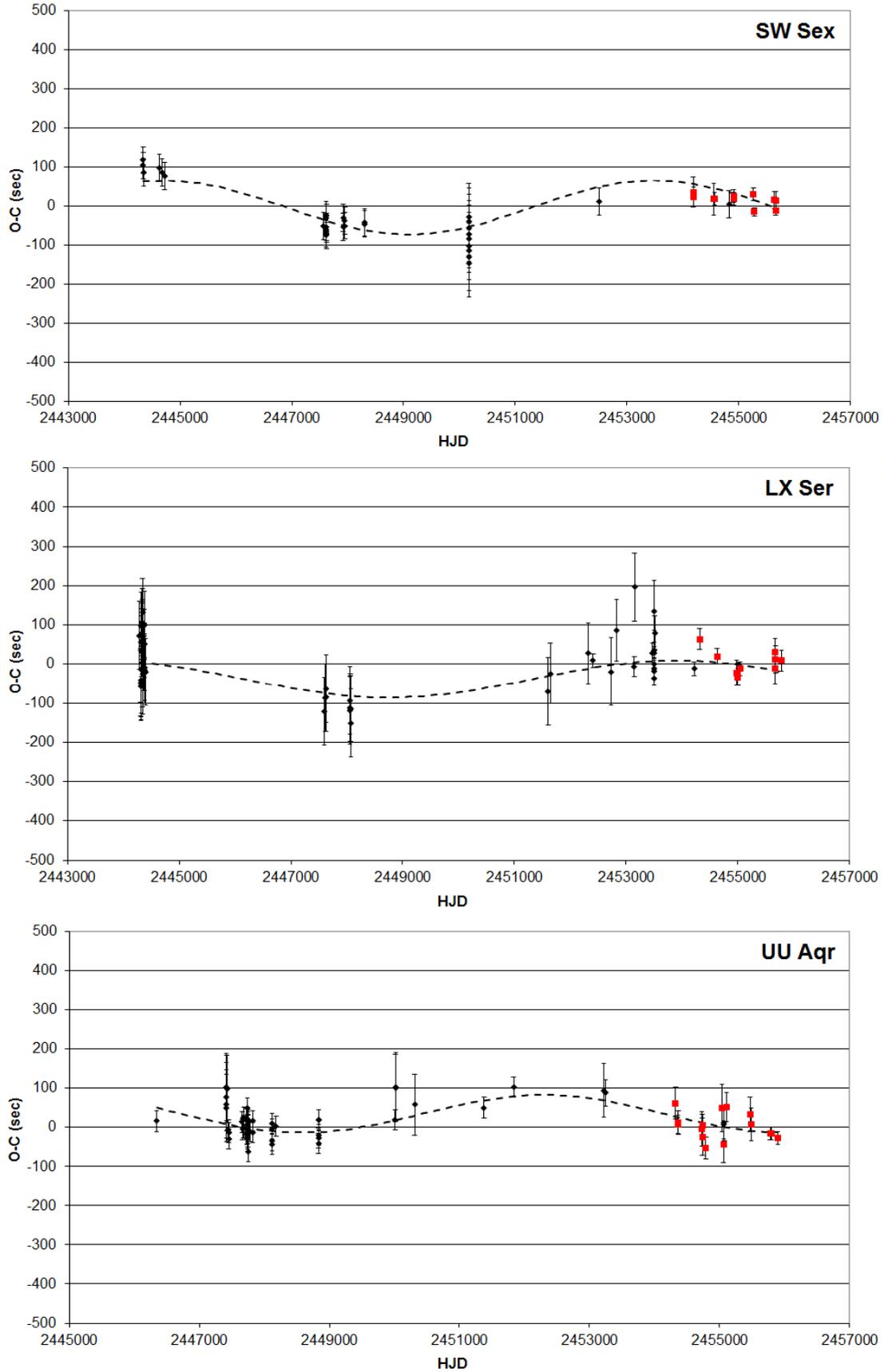

Figure 2. O–C diagrams with respect to the linear ephemerides in Table 3 for those SW Sex stars with Porb < 4 hrs which show possible cyclical variation in orbital period (dashed lines).

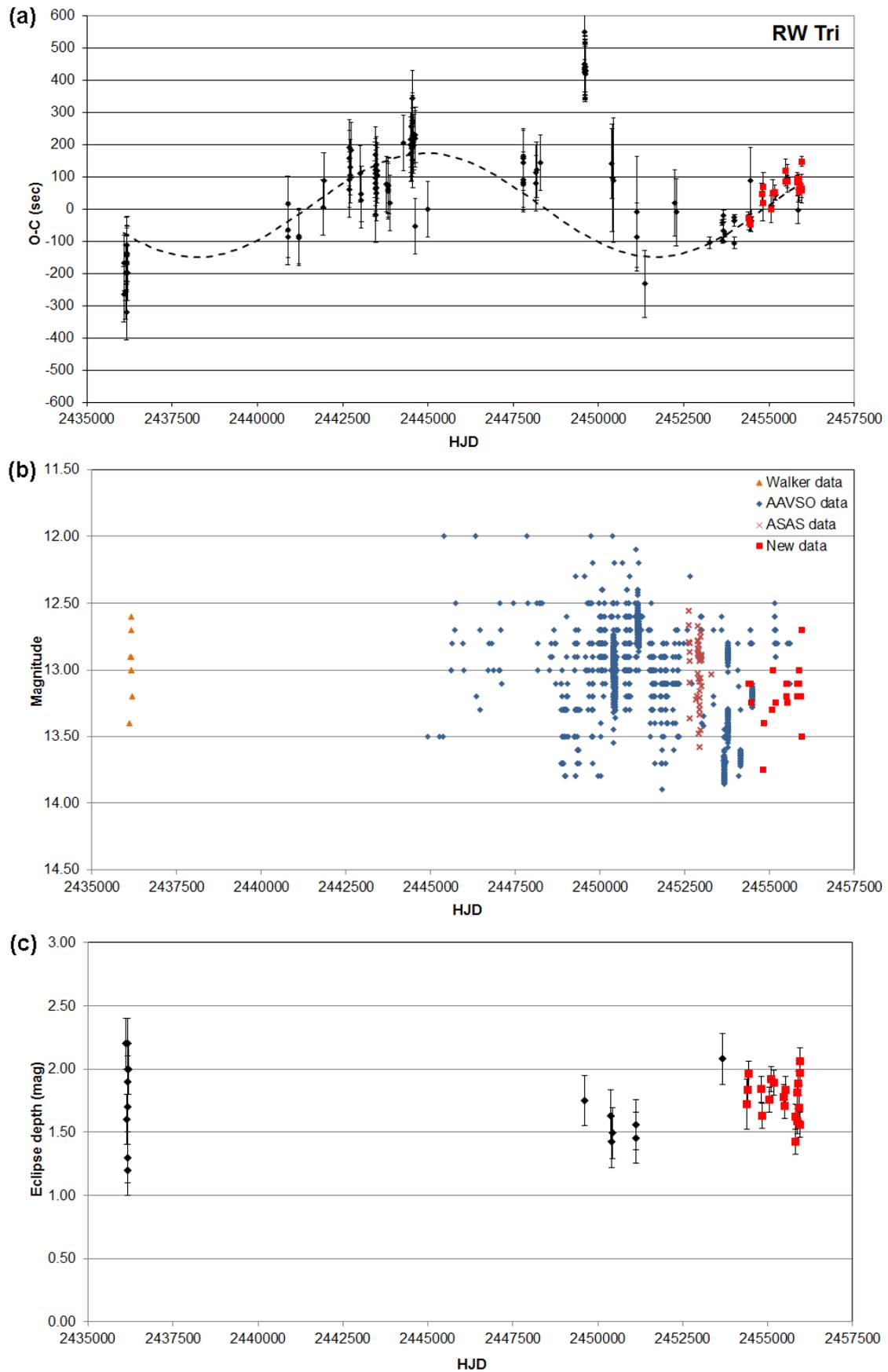

Figure 3. RW Tri: (a) O–C diagram with respect to a linear ephemeris showing a cyclical variation of orbital period (dashed line), (b) out-of-eclipse magnitude, and (c) eclipse depth.

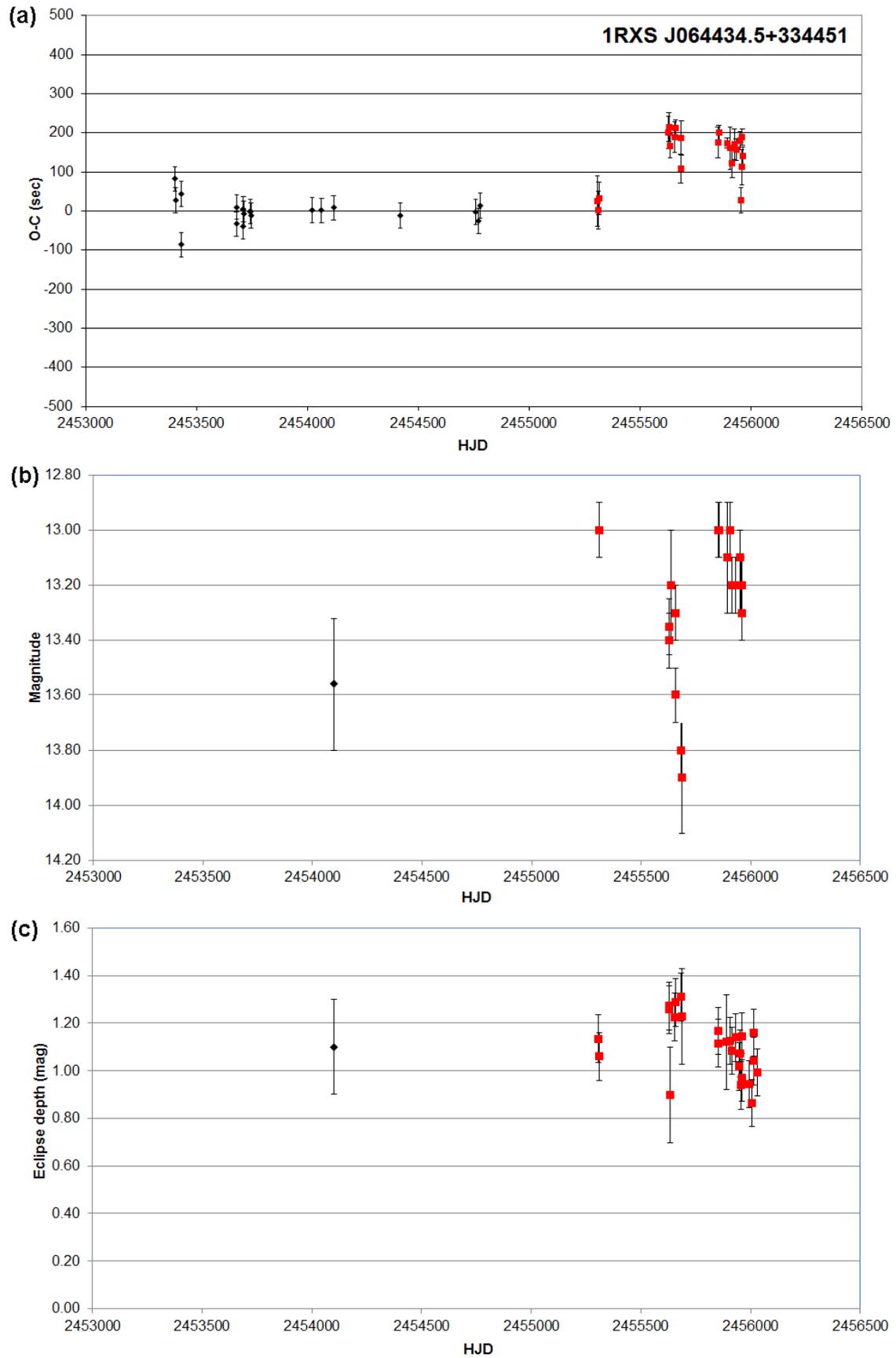

Figure 4. 1RXS J064434.5+334451: (a) O–C diagram with respect to a linear ephemeris for HJD < 2455500, (b) out-of-eclipse magnitude, and (c) eclipse depth.

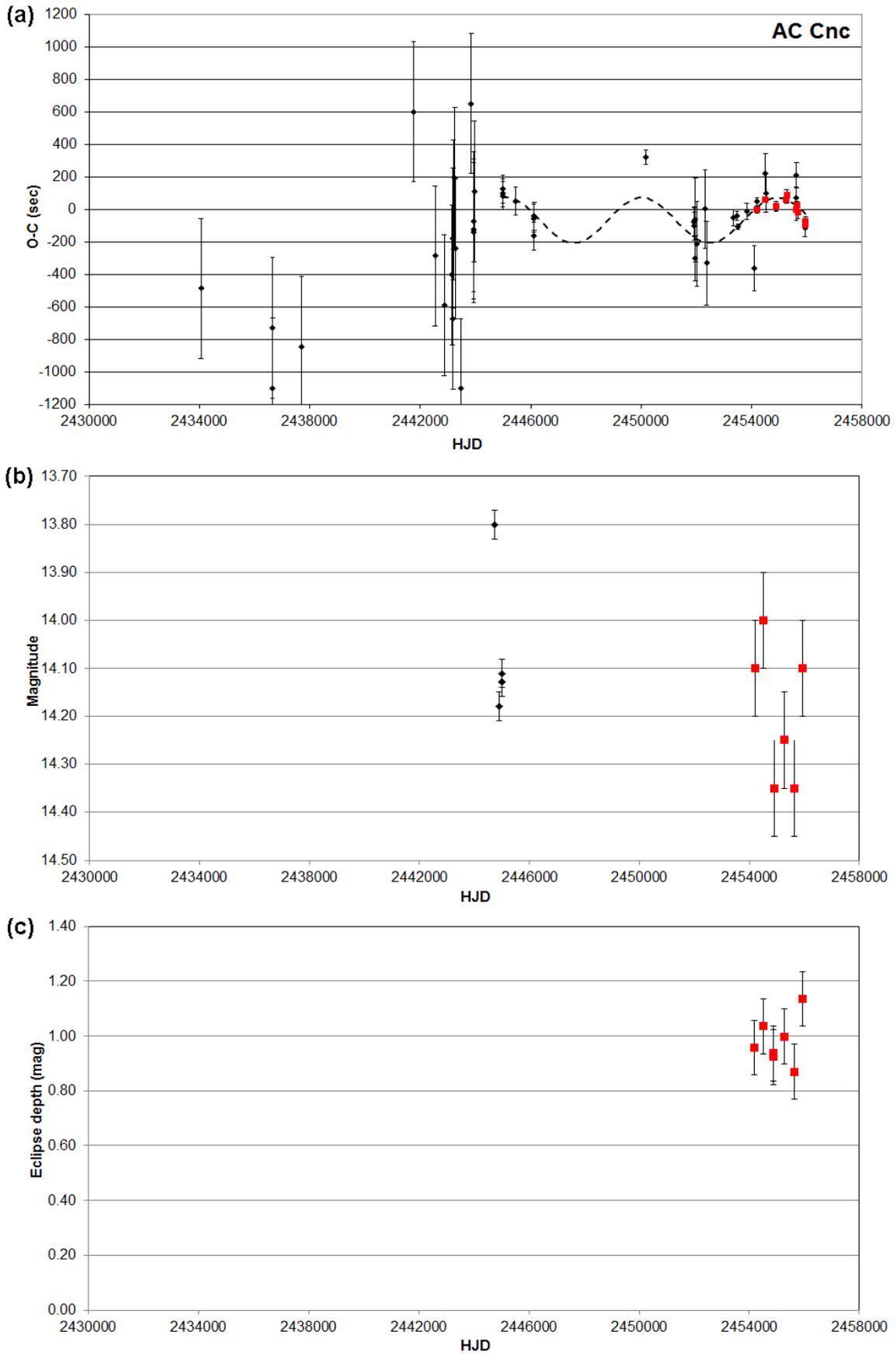

Figure 5. AC Cnc: (a) O–C diagram with respect to a linear ephemeris showing a cyclical variation of orbital period (dashed line), (b) out-of-eclipse magnitude, and (c) eclipse depth.

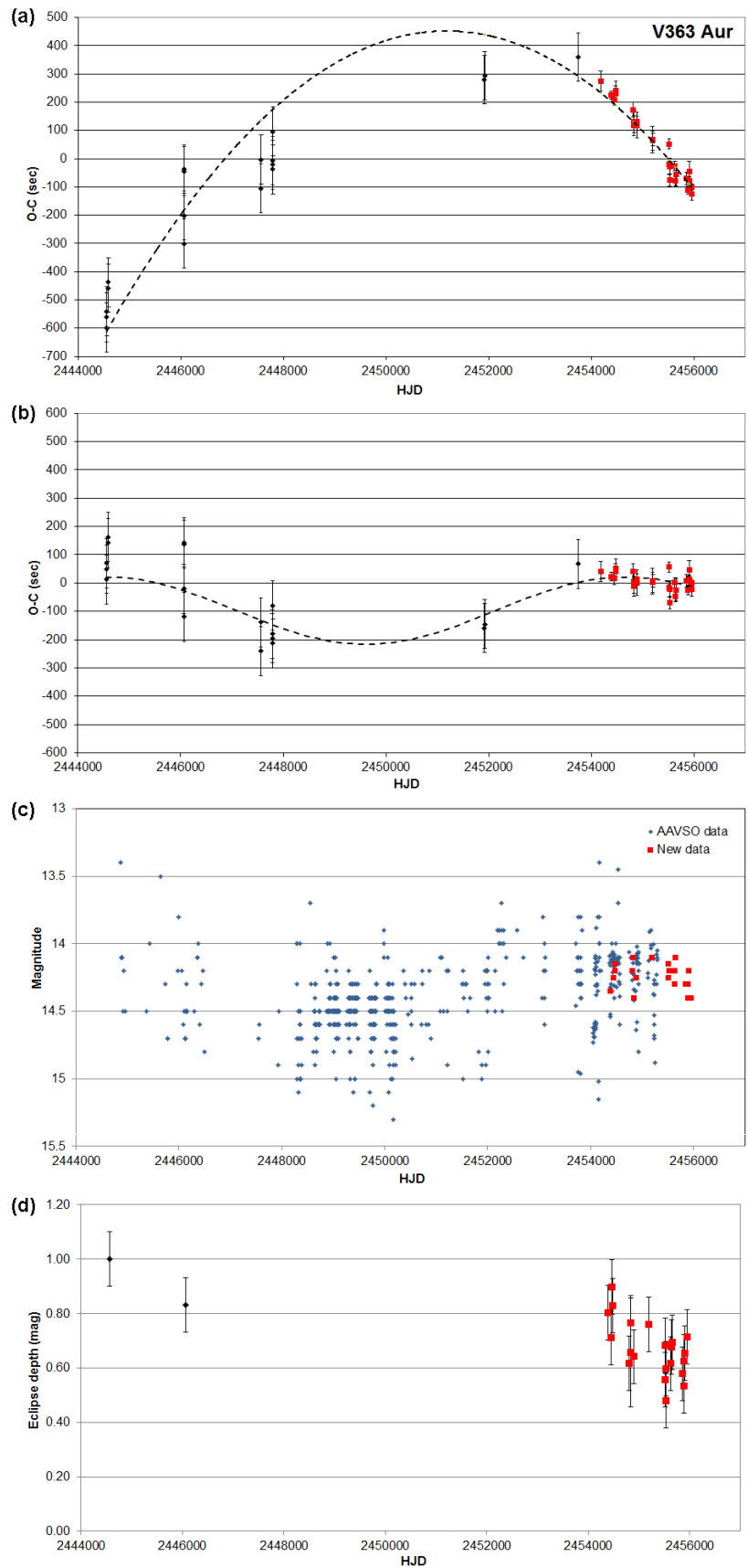

Figure 6. V363 Aur: (a) O–C diagram with respect to a linear ephemeris showing a quadratic ephemeris (dashed line), (b) O–C diagram with respect to a quadratic ephemeris showing a cyclical variation of orbital period (dashed line), (c) out-of-eclipse magnitude, and (d) eclipse depth.

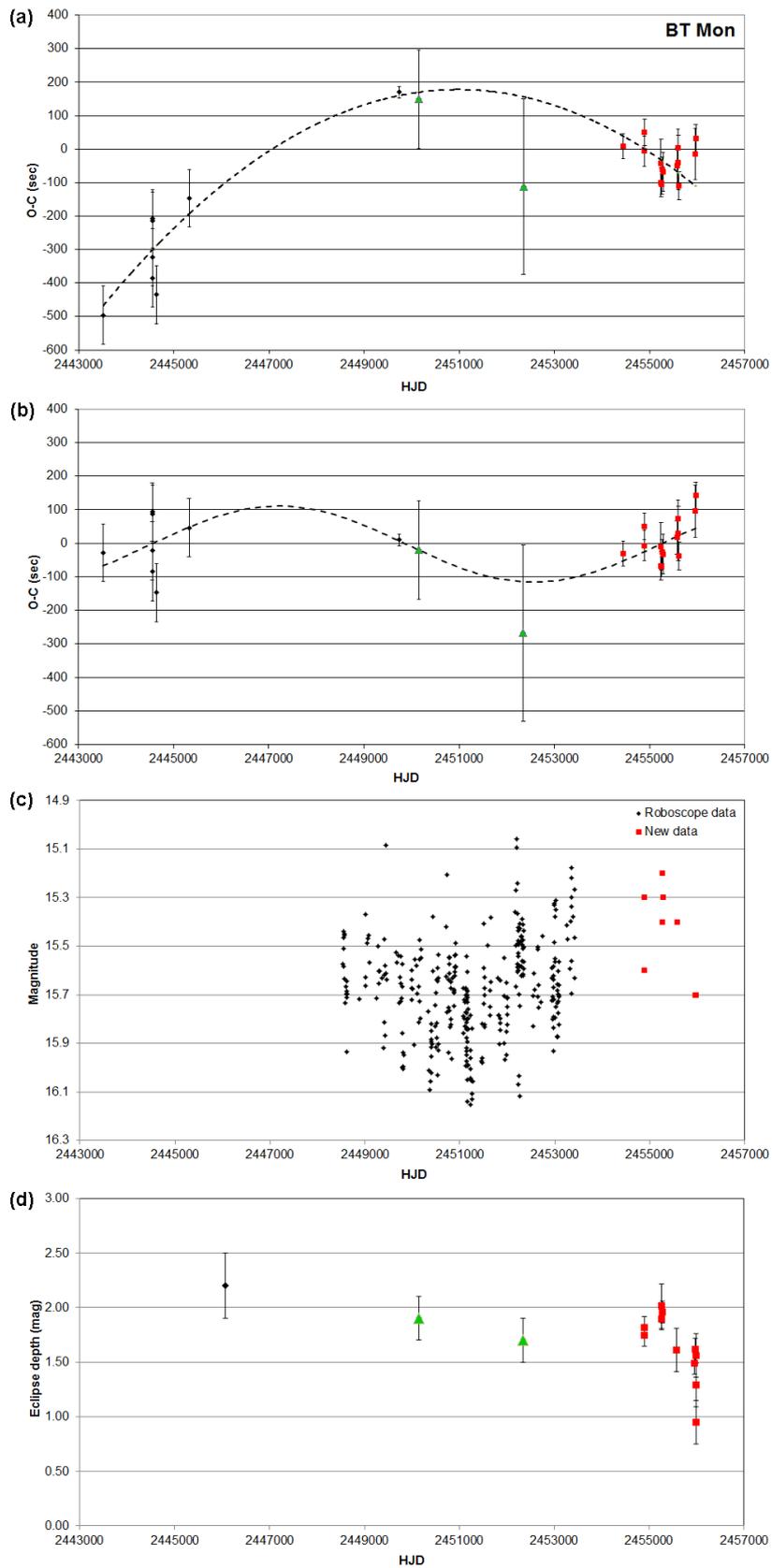

Figure 7. BT Mon: (a) O–C diagram with respect to a linear ephemeris showing a quadratic ephemeris (dashed line), (b) O–C diagram with respect to a quadratic ephemeris showing a cyclical variation of orbital period (dashed line), (c) out-of-eclipse magnitude, and (d) eclipse depth. Eclipses synthesised using Roboscope data are shown as green triangles.